\documentclass[preprint]{aastex}

\newcommand{\etal}{{\it et al.}\ }

\shorttitle{von Hippel \& Gilmore}
\shortauthors{NGC 2420 White Dwarfs}

\begin{document}

\title{The White Dwarf Cooling Age of the Open Cluster NGC 2420}

\author{Ted von Hippel}
\affil{Gemini Observatory, 670 North A'ohoku Place, Hilo, HI 96720, USA \\
email: ted@gemini.edu}

\author{Gerard Gilmore}
\affil{Institute of Astronomy, Madingley Road, Cambridge CB3 0HA, UK \\
email: gil@ast.cam.ac.uk}

\begin{abstract}

We have used deep HST WFPC2 observations of two fields in NGC 2420 to
produce a cluster CMD down to V $\approx 27$.  After imposing
morphological selection criteria we find eight candidate white dwarfs in
NGC 2420.  Our completeness estimates indicate that we have found the
terminus of the WD cooling sequence.  We argue that the cluster distance
modulus is likely to be close to $12.10$ with E(B$-$V) = $0.04$.  With
these parameters we find a white dwarf cooling age for NGC 2420 of $2.0
\pm 0.20$ ($1 \sigma$) Gyrs.  The $0.20$ Gyr uncertainty includes errors
in the photometry, sequence fitting, precursor time scales, and
theoretical white dwarf cooling time scales.  Comparing the cluster white
dwarf cooling age to ages derived from stellar isochrone fitting we find a
preference for ages derived from models incorporating convective
overshoot.

\end{abstract}

\keywords{Galaxy: stellar content -- open clusters and associations:
individual (NGC 2420) -- stars: evolution -- white dwarfs}

\section{Introduction}

NGC 2420 has been the subject of numerous investigations primarily due to
its combination of richness and age.  Richness, of course, allows one to
find stars in many of the rarer stages of stellar evolution, and in turn
make detailed comparisons between stellar evolution models and the
properties of either individual stars or the cluster as a whole via the
cluster color magnitude diagram (CMD) and luminosity function (LF).  NGC
2420 is roughly 2 Gyrs old, placing it on a logarithm age scale
approximately evenly between the ubiquitous young open clusters such as
the Hyades and the oldest star clusters known, the Galactic globular
clusters.  Overall, stellar evolutionary theory is in an advanced state
with sophisticated predictive abilities, including ages that are rapidly
becoming more reliable (e.g. Pols \etal 1998; Dominguez \etal 1999).  A
number of important uncertainties remain in stellar evolution theory,
however.  One of the two biggest uncertainties is the calibration between
the theoretical temperature - luminosity plane and the observational color
- magnitude plane.  The other of the two biggest uncertainties is the
theory of convection (note, however, the recent advances of Canuto and
coworkers, e.g. Canuto 1999; Canuto \& Dubovikov 1998), relevant both near
the surfaces of stars with T$_{eff} \leq 6500$ and in the cores of stars
more massive than the Sun.  Because of its age, and therefore the mass of
stars currently evolving off the main sequence, NGC 2420 provides an
important test of the degree of convection in stellar cores.  The details
of convection in stellar cores, in turn, have important ramifications
throughout astronomy.

Recent work in cosmology and galaxy evolution has led to increased
interest in stellar evolutionary ages.  For galaxy evolution studies much
current effort is being placed on determining the initial epoch of galaxy
building and star formation.  All of these studies rely on evolutionary
time scales set by stellar evolution.

Because of the importance of stellar ages we have sought to test stellar
evolution theory itself from outside the traditional approaches.
Traditional tests of stellar evolution are based on using stellar
evolution models to reproduce the properties of stars in stellar clusters
and binary pairs.  White dwarfs (WDs), on the other hand, can be used as
chronometers with almost complete independence from the theory of main
sequence stellar evolution.  There is a well-defined relation between the
luminosity and age of a white dwarf (e.g. Iben \& Tutukov 1984; Wood 1992;
Salaris \etal 1997), especially during the first few billion years of WD
cooling before crystalization effects become important.  White dwarfs have
been recognized as potential Galactic chronometers since at least the
proposal by Schmidt (1959) that the age of the Galactic disk could be
found via the luminosity limit of local WDs.  Subsequently a number of
studies (e.g. Winget \etal 1987; Liebert, Dahn \& Monet 1988; Wood 1992;
Oswalt \etal 1996; Leggett, Ruiz \& Bergeron 1998) of the WD LF have
measured the age of the Galactic disk.  White dwarfs have also been found
in clusters ranging in age from the Pleiades (one WD in this $\approx 70$
million year old cluster) to the globular clusters.  White dwarf cooling
ages have been derived for numerous young clusters, e.g. NGC 2451 (Koester
\& Reimers 1985), as well as a few clusters of intermediate age, e.g.
Praesepe (Claver 1995), NGC 2477 (von Hippel, Gilmore, \& Jones 1995,
hereafter paper 1), NGC 2420 (paper 1), and M 67 (Richer \etal 1998).  In
addition, lower limits that do not yet test stellar evolution theory have
been derived via WD cooling ages for NGC 188 (von Hippel \& Sarajedini
1998) and M 4 (Richer \etal 1997).  A number of other open and globular
clusters are also known to contain WDs.  For a recent summary of known
cluster WDs see the compilation of von Hippel (1998).

Overall, there is a clear consistency between the WD cooling ages and the
isochrone ages, which is comforting to note.  Within this overall
consistency, however, there are stellar evolution models with ages that
either do or do not closely match the WD cooling ages.  In this paper we
make a detailed comparison between the WD cooling age we derive for NGC
2420 and nine modern isochrone ages.  From this comparison we find that
models with convective core overshoot are favored over canonical (no
convective core overshoot) models for this cluster.

In order to avoid confusion, we note that some authors (e.g. Trimble \&
Leonard 1996; Dominguez \etal 1999) interpreted the WD ages we (von Hippel
\etal 1995) derived for NGC 2420 and NGC 2477 as inconsistent with any
isochrone ages for these clusters.  Certainly, there was no consistency
between the isochrone ages available at that time and our WD age for NGC
2477.  On the other hand, in 1995 there were no modern isochrone studies
for NGC 2477.  At that time, the most recently derived age for NGC 2477
(Carraro \& Chiosi 1994) was not based on isochrone fitting, but rather on
an isochrone-based calibration of the magnitude difference between the
main sequence turn off and red clump luminosity.  An improved WD age with
a more detailed analysis of NGC 2477 will be the subject of a future
study.  For the present study we focus on NGC 2420, where the isochrone
and WD age comparisons yielded ambiguous results in paper 1.  The primary
purpose of paper 1 was to show the power of using WD cooling time scales
as an independent test of ages derived from stellar evolutionary theory.
We will show here that the increased reliability of the observational
results for NGC 2420, as well as the large body of new theoretical work
for this cluster, makes it worthwhile to recompare the WD and isochrone
ages for this cluster.  Finally, we note that the current paper supersedes
paper 1 as it contains more and deeper observations obtained with an
improved observational technique, and it is based on a much-improved
understanding of the WFPC2 calibration.  Paper 1 was based on data taken
within three months of the installation of WFPC2, before calibrations were
well-established.

\section{Observations and Reductions}

We observed a single field in NGC 2420 with the Hubble Space Telescope
(HST) and the then new Wide Field Planetary Camera 2 (WFPC2) in cycle 4.
These observations were designed to find white dwarfs in NGC 2420 to the
terminus of the cooling sequence.  Limited experience with WFPC2 at that
time resulted in a data set in which hot and warm pixels were hard to
distinguish from faint white dwarf candidates.  Our study of the cluster
WDs and their implied age was presented in paper 1.  After our experience
with the cycle 4 data it became clear that a reliable determination of the
terminus of the WD cooling sequence required dithered observations as well
as twice the number of cluster stars.

In cycle 6 we obtained two additional V-band (F555W) pointings on our
cycle 4 field (hereafter, field 1), each slightly offset from the other
and from the cycle 4 pointing.  We also obtained new V- and I-band (F814W)
observations of a second field (hereafter, field 2).  We obtained a
slightly different set of individual exposure lengths and total exposure
times for field 2, based on our experience with field 1.  The details of
the observations are given in Table 1.  The first column of Table 1 lists
the dates of the nine different NGC 2420 pointings.  The second column
lists the number of exposures and the length of each exposure in seconds.
The last five entries indicate three exposures composed of individual
exposures of both $700$ and $800$ seconds.  The third column lists the
WFPC2 filter used, the fourth column lists the observed sky value in
counts, the fifth column lists the field identifier we use throughout the
text, and the sixth column lists the HST cycle number in which the
observations were made.  Fields 1 and 2 are centered $3.4$ arc minutes SW
and $1.4$ arc minutes NE of the apparent cluster center at 7:38:11.2,
$+$21:33:09.7 (J2000.0) and 7:38:28.5, $+$21:35:29.1 (J2000.0),
respectively.

We recalibrated the entire data set using the Canadian Astronomy Data
Centre's\footnote{CADC is operated by the Herzberg Institute of
Astrophysics, National Research Council of Canada.} archive pipeline with
up-to-date calibration files.  Following recalibration, we stacked the
aligned images of each subfield and rejected the cosmic rays with the
IRAF\footnote{IRAF is distributed by the National Optical Astronomy
Observatories, which are operated by the Association of Universities for
Research in Astronomy, Inc., under cooperative agreement with the National
Science Foundation.} task CRREJ.  Since there were only a few stars in the
PC frames we dropped the PC frames of both fields from all further
reduction and analysis.  We then used the drizzle package (Fruchter \&
Hook 1998) to shift and stack the dithered WF frames to arrive at a
combined V and a combined I frame for each of fields 1 and 2.  The drizzle
process rejected most hot pixels and other image defects and produced
reasonable stellar point spread functions (PSFs).  The PSFs were sometimes
a bit distorted, due probably to the fact that there were only two or
three pointings per filter per field.  The effect of the PSF shapes on the
quality of the photometry was small, as will be discussed below.  The
primary purpose for obtaining dithered observations and incorporating
drizzled reductions was to remove the hot pixels, which worked well.  We
employed SExtractor (Bertin \& Arnouts 1996) to find and classify sources
and CCDCAP\footnote{IRAF implementations of CCDCAP are available via the
web at the following site:~ http://www.noao.edu/staff/mighell/ccdcap/.}
(Mighell 1997), an aperture photometry task specifically designed for
WFPC2 data, to derive instrumental magnitudes.

As is well known by now, a number of small corrections must be applied to
WFPC2 photometry to fully remove instrumental artifacts.  We endeavored to
apply these corrections as best we could based on the current knowledge of
the WFPC2 instrument.  The effect of geometrical distortions on the WFPC2
photometry was corrected by drizzle.  To remove the effects of the charge
transfer (in)efficiency (CTE) problem we used the algorithms of Stetson
(1998) and Whitmore, Heyer, \& Casertano (1999).  The Whitmore \etal CTE
algorithm is independent of the so-called ``long versus short exposure
effect'', which we corrected using the prescription of Casertano \&
Mutchler (1998).  Both the Stetson (1998) and Whitmore \etal (1999) plus
Casertano \& Mutchler (1998) approaches gave typical corrections rising
from $\approx 0.02$ mag at V or I = $16$ to $\leq 0.10$ mag at V or I =
$26$.  While these two approaches differ in mathematical form and only the
latter corrects the time-dependent nature of the CTE, for our data they
both gave very similar results.  The mean V magnitude difference between
these two approaches among the faint (V = $22$ to $26$) stars is $\leq
0.01$ mag.  We take this small difference as an indication that we have
reliably corrected the effects of CTE, with an estimated uncertainty in
this correction no more than double the difference between these two
prescriptions, i.e. $\leq 0.02$ mag.

To determine the photometric drift of the WFPC2 with time we analyzed the
relevant calibration data (see Gonzaga \etal 1999) available via the Space
Telescope Science Institute's web pages.  For the epochs of our
observations the corrections were always $\leq 1$\%, except for WF3 at one
epoch (4/2/96), which required corrections of $+1.5$\% and $+2.0$\% in
F555W and F814W, respectively.  We chose only to correct these two cases
where the photometric drifts had exceeded $1$\%.  Since even these frames
were combined with other frames from epochs with essentially no
photometric drift, the resulting corrections were only $0.005$ and
$0.0067$ mag in F555W and F814W, respectively.  The error on these
corrections is likely to be $\leq 0.002$ mag.

The fairly large number of well exposed stars on each WF chip allowed us
to measure aperture corrections for each combination of WF chip, filter,
and field.  The aperture corrections had typical errors of $0.01$ to
$0.015$ mag.  We also investigated whether there were any
spatially-dependent aperture corrections and found none.  We did not apply
any breathing corrections to the photometry.  The fact that neither
spatially-dependent aperture corrections nor breathing corrections were
necessary was most likely due to the fact that each combined image
represented from four to ten exposures covering always more than one
orbit.  We also did not apply corrections for the photometric offsets due
to pixel size differences at every 34th row (Anderson \& King 1999).
Although this correction statistically affects $6$\% of the stars by
$0.01$ to $0.02$ mag, the manner in which we combined our photometry with
drizzle meant that the 34th row effect would induce errors in $18$\% of
our photometry, but only by $0.003$ to $0.007$ mag.

Finally, the data were transformed to the Johnson V and Kron-Cousins I
system via the equations of Holtzman \etal (1995).  While the F555W and
F814W filters transform well to the standard V- and I-band system, there
is naturally an error associated with this step as well, estimated to be
$\approx 2$\% (Holtzman \etal 1995).  The Holtzman \etal photometric
transformations are applicable over the color range $-0.3 <$ V$-$I $<
1.5$, whereas our stars continue to V$-$I $\approx 3.2$.  The emphasis in
this paper is on the cluster white dwarfs, which are within the color
limits of the Holtzman \etal transformations.

Our photometry is presented in Figure 1, where the error bars represent
only the internal, statistical photon-counting error.  The external
systematic error in the important cool WD region of the color magnitude
diagram, near V = $25.6$ and V$-$I = $0.6$, is the quadrature sum of the
above errors, and amounts to $\approx 0.03$ mag.  The dominant sources of
systematic error are the CTE corrections and the photometric
transformations.

\section{Discussion}

\subsection{The Color Magnitude Diagram}

There are a large number of galaxies and remaining image defects with the
approximate color and magnitude of the faint WDs in NGC 2420 (Figure 1).
One of the major motivations for using HST for this study is the ability
of this telescope to resolve nearly every galaxy in the Universe, when
sufficient signal-to-noise is obtained.  Thus, if image defects can be
eliminated any unresolved object is almost surely a star.  The converse of
this is also true, any resolved object, even marginally resolved, is not a
star.  For our purposes, we required good galaxy and image defect
rejection beyond the limit of the WD cooling terminus, at V $\approx
25.6$.  Our data are of sufficient quality to do this, and the SExtractor
classifications provided an easy means of making this separation.

Figure 2 shows the results of the SExtractor morphological classification
versus V-band magnitude.  The ``stellarity index'' ranges from $0$
(galaxies) to $1$ (stars).  Careful examination of the images revealed
that all objects with stellarity index $\geq 0.9$ are unresolved (i.e.
stars).  Objects with stellarity indices between $0.9$ and $0.6$ are a
mixture of unresolved objects, very faint objects, and image defects.  All
objects with stellarity index $\leq 0.2$ are resolved, although some are
not galaxies, but rather ghost images or part of a diffraction spike.
Objects with stellarity indices between $0.5$ and $0.2$ are a mixture of
resolved objects, very faint objects, and image defects.  The term image
defects here is meant to include the remaining hot and warm pixels, as
well as residual remaining cosmic rays, ghost images, diffraction spikes,
etc.

The large number of definite stars near the top of Figure 2 demonstrates
that a star cluster is present in this field.  Contamination by background
galaxies becomes significant at V $\approx 25$ and near V = $27$ a
combination of objects too faint to reliably classify and image defects
predominate.  Note also that saturated stars have a stellarity index of
somewhat less than $1.0$ due to their flat-topped, broader PSFs.  The
stellarity index cut of $0.78$ is drawn in Figure 2.  This choice was
somewhat relaxed from the classification value for a typical well-exposed
star since the drizzle processing created some slightly distorted PSFs.
To be considered a star, an object had to have a stellarity index of $\geq
0.78$ in either the V- or the I-band frames.  By allowing the
classification to be based on either frame, we were able to bypass some of
the reduced classification probabilities due to the distorted PSFs caused
by the drizzle process, as well as take into account that some of the
faintest stars are better observed in one filter than the other, depending
on their color.  Nonetheless, we tried to be conservative for each of the
stars that matter in this study, i.e. the WDs, and we additionally
examined each of these detections by eye to check for adjacent image
defects, crowding with stars or galaxies, or any other problem that might
compromise the morphological classifications or the photometry.  The
subject of morphological classification and how one compromises between
ensuring that only stars are counted and ensuring that no stars are missed
will be revisited in Section 3.3.  For now we take all objects with a
stellarity index $\geq 0.78$ as stars and plot them in Figure 3.

The reader might wonder whether proper motions could be used in this
cluster to help differentiate members from non-members.  While it is true
that our field 1 observations span a period of two years, all of our field
2 observations were obtained on the same day (see Table 1).  Furthermore,
the proper motion of the cluster is known to differ from the mean proper
motion of bright (B $\leq 13.5$) stars in this field by $\approx 0.002$
arcsec yr$^{-1}$ (van Altena \& Jones 1970), or only $0.04$ WF pixels
over a two year baseline.  The primary difference between the cluster and
field is in the dispersion of proper motions, with measured values
corresponding to $0.014$ and $0.064$ WF pixels over the two year baseline
for the cluster and field stars, respectively.  We were thus not surprised
when we were unable to detect a difference between cluster and field stars
in field 1 based on the change in image centroids.  Our typical measuring
error appeared to be $\approx 0.1$ pixel.  These errors could perhaps be
reduced by deriving optimized centroids, but we judged the likely
additional information to be minimal.  Fortunately, proper motion
information is not required as there are few remaining contaminating
objects, either Galactic field stars or unresolved background galaxies, in
the WD portion of the CMD (Figure 3).

The CMD presented in Figure 3 shows a clear main sequence extending $10$
magnitudes from V $\approx 16$ to V $\approx 26$.  The apparent gap in the
main sequence from V $\approx 19$ to $21$ is likely just a statistical
variation in the cluster luminosity function for our two fields, visually
exaggerated by the pile up of saturated photometry near V = $16$.  Wider
field coverage of NGC 2420 would demonstrate whether this gap is a real
and unexpected variation in the cluster LF.  Note that the ground-based
CCD photometry of Anthony-Twarog \etal (1990), which covered $20$ times as
much of the cluster as our study, does not show this gap, though part of
the gap is beyond the limit of their photometry.  Regardless of the
astrophysical meaning of the gap, it is not caused by any form of
incompleteness in the HST photometry.  Even stars one magnitude fainter
than the bottom of the gap are readily visible to the naked eye in the raw
uncombined images even before cosmic ray rejection.

The two separate clumps of saturated stars are the result of our use of
exposure times ranging from $700$ to $1200$ seconds in F555W.  The cluster
main sequence binaries are also visible, particularly between V = $21$ and
$24$.  In the lower left of the CMD a series of blue stars from V $\approx
21.4$ to $25.6$ closely follow the cooling track for a $0.7$ solar mass
Carbon-Oxygen WD model (Benvenuto \& Althaus 1999).  The WD model was
placed at a distance modulus of $12.10$ (see Section 3.2).  There are also
a few dozen stars sprinkled throughout the CMD between the WD sequence and
main sequence.  These are Galactic field stars behind the cluster (see the
discussion at the end of Section 3.1).

As a final tool to understanding objects in the observed CMD (Figure 3) we
employed the Galaxy model of Reid \& Majewski (1993) to create a model CMD
for Galactic field stars at the location of NGC 2420.  This model has been
tested against north Galactic pole number counts and color distributions
(Reid \& Majewski 1993) and against two deep, lower latitude fields (Reid
\etal 1996).  The model CMD is presented in Figure 4.  Note that the model
predicts that there should be some Galactic field stars near the cluster
main sequence and a few more sprinkled in the region between the main
sequence and the WD sequence, roughly as seen in the observed CMD.  By
increasing the model normalization by a factor of $10$ we find the model
predicts that $0.6$ Galactic disk WDs lie somewhere along the observed
cluster WD sequence.  This implies that there may be a single interloper
somewhere along the cluster WD cooling sequence.  The likelihood of a
Galactic WD interloper in the magnitude beyond the observed WD cooling
sequence limit is much lower, however, as only $0.2$ WD interlopers are
expected in this region.  We do not want to over-interpret this
model-dependent estimation of the numbers of WD interlopers, however, i.e.
by claiming that a $+4 \sigma$ enhancement in Galactic WD numbers would be
required in this field to contaminate the WD terminus, since the Reid \&
Majewski model has never been tested on faint field WDs.  We are simply
using these model predictions to argue that it is unlikely that a Galactic
field WD is among the faintest observed WDs in our CMD.  We further note
that such an interloper need not affect the derived cluster WD cooling age
if the cluster age is determined from WD isochrone fitting, rather than
from just the faintest WD (see Section 3.3).

\subsection{Cluster Parameters}

The most commonly used values for the distance modulus and reddening for
NGC 2420 are (m$-$M)$_{\rm V} = 11.95$ and E(B$-$V) = $0.05$, values
largely supported by the data and analysis of Anthony-Twarog \etal
(1990).  Anthony-Twarog \etal fit VandenBerg (1985) isochrones to their
photometry.  Are results derived from these older models still the most
reliable?  Indeed, Anthony-Twarog \etal noted the poor fit between the
cluster turn-off region and the VandenBerg (1985) isochrones.  Subsequent
reanalysis of the Anthony-Twarog \etal data by Demarque, Sarajedini, \&
Guo (1994), incorporating up-to-date stellar evolution models, yielded
(m$-$M)$_{\rm V} = 12.05 \pm 0.10$ and E(B$-$V) =
$0.045^{+0.020}_{-0.015}$.  In another reanalysis, Twarog, Anthony-Twarog,
\& Bricker (1999), in a detailed study of the red giant clump luminosity
in NGC 2420 and other open clusters, employed stellar evolution models and
main sequence fitting to rederive the parameters for NGC 2420.  Twarog
\etal argued that (m$-$M)$_{\rm V} = 12.15$ and E(B$-$V) = $0.04$.  We
take these two analyses as being the most up-to-date and conclusive on the
issue of the cluster distance and reddening, and adopt the mean of their
distance moduli, $12.10$.  The reddening values for the two studies are
entirely consistent, and we adopt E(B$-$V) = $0.04$.\footnote{In paper 1
we adopted (m$-$M)$_{\rm V} = 11.95$ and E(B$-$V) = $0.05$.}  In the
analysis of the WD cooling ages, below, we will determine WD age as a
function of assumed distance modulus since a wide range of distance moduli
have been used in the isochrone fits.  The value of the reddening does not
affect the cluster WD ages since the WD isochrone fits depend almost
entirely on V luminosity, and not color.  It is comforting to note,
however, that the cluster redding is low and consistent in these modern
studies.

We derived an independent distance modulus by fitting the open cluster
fiducial main sequence presented by Pinsonneault \etal (1998) to the
Anthony-Twarog \etal data.  We found a distance modulus of $11.85$ to
$12.02$ for [Fe/H] = $-0.4$ and $11.95$ to $12.10$ for [Fe/H] = $-0.3$.
Our distance modulus is consistent with that of Twarog \etal (1999) given
their assumed metallicity, [Fe/H] = $-0.29$.  Rather than using the
distance modulus we obtained from main sequence fitting, we rely on the
above average distance modulus of $12.10$, since the main sequence fitting
technique is so sensitive to the cluster metallicity.

\subsection{White Dwarfs}

In paper 1 (see Figure 2 of that study) we found five WD candidates in the
single HST pointing.  We recovered all five of these WD candidates in our
present analysis but only three of them (and only one of the faintest
three from paper 1) passed our morphological classification.  We imposed
tighter constraints in this study since we have both more data and a
greater knowledge of our data.  The two objects excluded may very well be
cluster WDs and may just have slightly deviant PSFs due to the drizzle
processing.  Regardless of the cause, we wanted to minimize suspect
objects, and so do not include these objects in the CMDs of Figures 2 and
3.  We note, however, that the V-band luminosities for these two rejected
potential WDs (V = $25.28$ and $25.16$ in our present data, V = $25.11$
and $24.87$ in our cycle 4 reductions) are $0.4$ mags brighter than the
faintest WDs presented here.  Their inclusion would not affect our derived
WD age.  Our candidate WDs are listed in Table 2.  Column 1 lists a WD
identification number, ordered by brightness in the V-band.  Column 2
lists the object V magnitude, followed by its uncertainty in column 3.
Column 4 lists the V$-$I color, followed by its uncertainty in column 5.
Columns 6 and 7 list the object Right Ascension and Declination (J2000.0),
respectively.  The errors in position are expected to be $\approx 0.5$
arcsec.  The relative positions of objects in the same WFPC2 field should
be significantly more accurate, $\approx 0.1$ arcsec.

Of the eight candidate WDs presented in Table 2 and Figures 2 and 3, one
candidate (WD4) had a significantly lower probability ($0.71$ in V, $0.23$
in I compared with our threshold value of $0.78$) of being stellar,
according to the SExtractor stellarity index.  We nonetheless retain this
object since we believe its morphological classification appears
non-stellar due to crowding by an adjacent galaxy.  Furthermore, upon
carefully examining this object in the V- and I-band images it appeared
that the crowding galaxy had a similar color and is unlikely to greatly
change the derived color of this candidate WD.  Since we believe it is a
WD and since the hotter WDs hold none of the weight in the age fit, we
present it in our CMDs.

Before applying WD isochrones to our data, we first discuss the inputs
required for the WD isochrones.  First, we require a cluster distance
modulus to convert apparent magnitudes to absolute magnitudes.  Second, we
require the evolutionary ages of the precursor stars, including the time
required for them to evolve from the main sequence, through the giant
branch, through any subsequent burning stages, and through the planetary
nebula stage, until they become WDs.  For precursor ages we rely on the
stellar evolution parameterizations of Hurley, Pols \& Tout (2000).  To
connect the precursor masses to the WD masses, we employ the initial -
final mass relation of Wood (1992).  Other modern studies (e.g. Koester \&
Reimers 1996; Dominguez \etal 1999) of the initial - final mass relation
are consistent at the level required for our purposes.  

It may seem counterintuitive that ages derived via WD luminosities could
be independent of stellar evolution theory since the total age of the WD
depends on the precursor ages, but the rapid evolution of high mass stars
means that the precursor time scales have little leverage on the total
age.  The WDs that are presently the coolest and faintest in any cluster
are those which formed first, and therefore those which evolved from the
most massive progenitors.  The first stars to become WDs had a main
sequence mass somewhere between $6$ and $8$ solar masses with total
evolutionary time scales of $\leq 8 \times 10^7$ yrs, i.e. $\leq 5$\% of
the $\approx 2$ Gyr ages considered here.  In order to quantify the
uncertainty in the WD ages due to the uncertainty in the precursor
lifetimes we measured the difference in total age after adjusted the
precursor masses by $\pm 20$\%.  This change in precursor mass should
account for both uncertainty in the initial - final mass relation and
uncertainty in the evolutionary time scales themselves.  For example, the
Wood initial - final mass relation gives a precursor of $3.67$ solar
masses for a $0.7$ solar mass WD.  The $\pm 20$\% mass values become
$4.40$ and $3.06$ solar masses and the evolutionary time scales of these
stars are $0.160$ and $0.412$ Gyrs (Hurley \etal 2000).  For a $0.9$ solar
mass WD the $\pm 20$\% precursor mass values correspond to evolutionary
time scales of $0.048$ to $0.104$ Gyrs.  These evolutionary time scales
vary by a factor of more than two, yet the effect of these precursor time
scale changes, weighted among WDs of different masses, results in a
cluster WD age uncertainty of only $-0.05, +0.07$ Gyrs.  Clearly, for
clusters of $\approx 2$ Gyr, realistic uncertainties in precursor time
scales are unimportant, and thus the cluster WD age is essentially
independent of stellar evolution theory.

As a check on the reliability of the WD cooling theory itself, we compared
the WD cooling models of different groups.  Since the Wood (1992) models
were provided to us in the form of WD isochrones (Ahrens 1999) we use them
as our fiducial set.  The Benvenuto \& Althaus (1999) and Hansen (1999)
models are in the form of cooling tracks for WDs of different masses.  To
these cooling models we added the precursor time scales from Hurley \etal
(2000), as discussed above.  Within the age range of $1.5$ to $2.5$ Gyrs,
as given by the Ahrens isochrones, we found that the Benvenuto \& Althaus
models were systematically older than the Ahrens isochrones by $0.18$ Gyr,
and the Hansen models were systematically older than the Ahrens isochrones
by $0.05$ Gyrs.  Rather than use an average result from the three
different sets of models, we use the Ahrens isochrones to derive our
cluster WD age and use the differences between the models as an indication
of the uncertainties in the WD cooling time scales.  We estimate the age
uncertainty to be $\approx 0.15$ Gyrs within the theoretical models
themselves, for clusters between $1.5$ and $2.5$ Gyrs.

Although we believe the morphology cut we have chosen properly separates
stars from galaxies, we now demonstrate the insensitivity of our results
to the chosen morphological cut.  The general issue is somewhat
complicated and depends on the goal of the particular study.  If we wished
to derive a minimum cluster age then we would need to reject every
possible galaxy and use only objects that are highly likely to be white
dwarfs.  On the other hand, if we wished to derive a maximum cluster age
then we would need to include any object that might be a cluster white
dwarf.  Even without the image morphology information, the shape of the WD
LF provides additional guidance, and as our analysis will show, the
minimum and maximum WD age are one and the same for these data.

Figure 5 presents all possible candidate WDs in our two fields.  These
objects are selected based on their photometric proximity ($1.5 \sigma$)
to the model cooling tracks for $0.6$ to $1.0$ solar mass Carbon-Oxygen
WDs (Benvenuto \& Althaus 1999) placed at the cluster distance.  The
symbols indicating the range of the stellarity index value, with the
filled symbols indicating the most reliable WD candidates.  Recall that
objects with stellarity index $\leq 0.2$ are obviously resolved to the
eye, and so we do not include them in Figure 5.  In fact, objects within
the $0.2 \leq$ stellarity index $\leq 0.5$ range are almost surely all
galaxies or image defects, but we include them in our analysis anyway, in
order to demonstrate that this form of noise will not approximate the
shape of a WD LF.  The dotted lines in the lower left of Figure 5 are the
probabilities of finding objects as a function of luminosity, with the
first and second numbers indicating completeness in fields 1 and 2,
respectively.  Since the completeness levels are different in fields 1 and
2, each object is labeled with the field in which it was found.

The completeness estimates presented in Figure 5 were determined by
scaling artificial star tests we performed on our cycle 4 data in paper
1 using the Tiny Tim package (Krist 1995).  Reproducing artificial star
tests for our combined, dithered and drizzled, cycles 4 plus 6
observations would have been laborious, but fortunately was unnecessary.
The cycles 4 and 6 observations were obtained with the WFPC2 in exactly
the same configuration.  Additionally, the faintest stars, i.e.  those for
which the issue of completeness is the most relevant, are faint enough
that sky noise dominates shot noise in the object by a factor of two and
read noise by a factor of four to six.  We were thus able to scale our
earlier completeness simulations to our combined cycle 4 plus 6 data set
based on the new cumulative exposure times and sky values.  We note that
the limiting depth of our CMDs is not dictated by the ability of the
software to find objects at a limiting flux level, but rather by the need
for sufficient signal-to-noise to obtain reliable morphological
classification and a photometric precision of $\leq 0.15$ mag in both V
and I.  As it turns out, both the needed photometric and morphological
precision lead to essentially the same limiting magnitude.  These
photometric and morphological precision cuts are the reason so few objects
are seen in Figure 3 fainter than V = $26$, even though Figure 2 shows
many objects detected at fainter magnitudes.

Figures 6a through 6f presents the LFs extracted by lowering the threshold
through each stellarity index cut-off for objects in field 2, with the
cut-off values indicated in each panel.  The bin widths are $0.25$ mag to
preserve the quality of the photometry and the apparent pile-up of objects
near V = $26.6$.  Only objects in field 2 are presented in Figures 6a-f as
the field 2 observations probe $\approx 0.5$ mags fainter than those of
field 1.  The cross-hatched histogram presents the observed luminosity
functions, whereas the unfilled histogram presents the luminosity
functions corrected for completeness.  The final, corrected LF bin of each
of panels c through f contains $200$ objects.  These panels are not
rescaled to view this final bin as the rest of the LFs would be
invisible.  Overplotted on each panel is the WD LF derived by Richer \etal
(1998) for the $\approx 4$ Gyr open cluster M67.  The Richer \etal WD LF
is the best open cluster, i.e. single-age burst, WD LF currently
available.  They presented their WD LF with bin widths of $0.5$ mag.
Their error bars are due to both counting statistics and background
subtraction errors.  We normalized and slid in V magnitude the Richer
\etal WD LF to match the identified LF peak.  While all the panels of
Figure 6 suffer from low number statistics, only Figures 6a and 6b display
reasonable LFs.  If one were to insist that Figures 6e or 6f contained
reasonable WD LFs, two new problems would emerge.  First, there should be
at least half a dozen observed candidate WDs in the quarter magnitude bin
just beyond the identified WD LF terminus.  Perhaps this is simply due to
low number statistics?  Second, the LFs of Figure 6e and 6f would imply
twice as many cluster WDs as main sequence stars, which would be more than
an order of magnitude more WDs than seen in any other star cluster or in
the solar neighborhood (von Hippel 1998).

In summary, though we find only four objects that we identify with the WD
cooling sequence terminus, we believe our identification is sound since 1)
we believe our morphological selection criteria are reasonable, 2)
relaxing those morphological criteria from $0.78$ to $0.7$ or $0.6$ does
not change the WD LF, and 3) even relaxing the morphological criteria to
extreme levels only creates LFs too absurd to be true cluster WD LFs.

A separate question is whether the cluster could have dynamically ejected
its faintest WDs.  Significant dynamical ejection of the oldest WDs is not
expected, however, since these WDs have higher masses than both the
younger WDs and the bulk of the main sequence stars.  Strictly speaking,
the ages derived from the WD terminus provide not a cluster age, but
rather a firm lower limit to the cluster age, since both photometric
incompleteness and stellar ejection could rob the cluster CMD of its
oldest WDs.  If stars appear to pile up at an observed WD cooling sequence
terminus, however, it is likely that the cluster age is equal to, or just
slightly greater than, the age implied by the faintest cluster WDs.

In Figure 7a we present the CMD of Figure 3 with distance and reddening
removed.\footnote{E(V$-$I) = 1.21 E(B$-$V) for the HST F814W filter, based
on equations 3a and 3b of Cardelli, Clayton, \& Mathis (1989).}  The white
dwarf region of the CMD is presented in Figure 7b.  Plotted in Figures 7a
and 7b are the Ahrens isochrones abutted to a $0.6$ solar mass C-O WD
cooling track (Wood 1992) for $0.5, 1.0, 1.5, 2.0, 2.5$, and $3.0$ Gyrs.
The blue hook at the bottom of each isochrone and among the coolest WD
candidates is the expected result of both the fact that cooling is a
function of WD mass and the fact that more massive WDs are bluer.  There
appears to be a color offset between the reddest WD candidate and the
Ahrens isochrones of $\approx 0.2$ mag, though this offset depends on the
assumed cluster distance.  Likewise, the brightest and bluest WD is
$\approx 0.03$ mag redder than the Ahrens isochrones.  The Ahrens
isochrones differ by a few hundredths of a mag in V $-$ I color from the
cooling sequences of Benvenuto \& Althaus (1999) and Hansen (1999); e.g.,
compare the location of these objects to the WD tracks of both Figure 5
and Figure 7b.  The most likely explanation is that the imperfect color
match between the models and candidate WDs is due to a simple combination
of random and systematic photometric errors along with small uncertainties
in the cooling track colors.  Fortunately, the WD isochrone fit depends
primarily on the theoretically and observationally more precise WD
luminosities, and only secondarily on the colors.

We find a best fit age of $2.0 \pm 0.1$ Gyrs, not including photometric
errors or errors in the distance modulus.  The photometric errors for the
faintest WDs range from $0.032$ to $0.039$, with a further systematic
uncertainty in the calibration of $\approx 0.03$ mag.  Since all four of
the faintest WDs contribute to the age derivation the total photometric
uncertainty is $\approx 0.05$ mag, corresponding to an age uncertainty of
$\approx 0.06$ Gyrs.  Combining the photometric, fitting, precursor, and
theoretical errors in quadrature we arrive at a best fit cluster age of
$2.0 \pm 0.20$ ($1 \sigma$) Gyrs, for (m$-$M)$_{\rm V} = 12.10$ and
E(B$-$V) = $0.04$.  We do not consider isochrone fits for WDs with Helium
atmospheres since most WDs in this luminosity range have Hydrogen
atmospheres.  We also do not consider errors in the distance modulus in
deriving a best age for NGC 2420 since it would be unfair to compare a WD
age derived with a certain distance modulus with the various isochrone
fitting studies which have assumed different distance moduli.  In our
comparisons between the cluster WD and main sequence ages in the next
section we instead derive a best fit age for a range of cluster distance
moduli.

\subsection{Isochrone Ages}

Determining a best value for the cluster isochrone age is complex.
Indeed, the complexity and importance of the age question is the reason
why we have chosen to apply the WD cooling age technique to this cluster.
Because of its age, NGC 2420 is an excellent candidate to test the reality
and, if real, the amount of convective core overshoot in stars of
intermediate stellar mass.

In the last decade, numerous studies have addressed the question of
convective core overshoot for NGC 2420 and derived ages with or without
this component in their models.  Earlier studies of NGC 2420 predate the
entire question of convective core overshoot.  Since our goal here is to
compare our WD ages with those stellar evolution ages that are in current
use, and in particular to address the question of convective core
overshoot, we consider the isochrone ages derived only over the last
decade.  Table 3 summarizes the isochrone fits to NGC 2420 that meet our
criteria, in chronological order first for the studies employing canonical
models, then for the studies employing convective core overshoot.  Column
1 lists the derived cluster age in Gyrs with any reported age uncertainty
in parentheses.  Column 2 lists whether the stellar evolution model
incorporated core convective overshoot or not.  Columns 3 and 4 list the
distance modulus and reddening, respectively.  We do not tabulate the
uncertainties in distance moduli and reddening since many authors did not
report these uncertainties.  Furthermore, the relevant issue for our study
is to know what distance modulus corresponds to the reported age so that
we can make the proper comparison between the WD and isochrone age fits.
Column 5 indicates whether the distance modulus and reddening were derived
(``D'') along with the cluster age or adopted (``A'') from other studies.
The relevance of derived versus adopted cluster parameters is that the use
of a highly improbably distance modulus or reddening may be an independent
indication of a problem with the isochrone fitting.  Column 6 lists the
metallicity used in creating the isochrones.  Since the metallicity of NGC
2420 is almost surely within or very close to the range $-0.30$ to $-0.40$
(Friel \& Janes, 1993), this column helps to identify where inappropriate
stellar evolutionary models may have been applied.  Column 7 lists the
reference(s).  Column 8 lists the WD age that would be derived for the
distance moduli and reddening listed in columns 3 and 4.  These WD ages
are not preferred in any way, but rather are meant to serve as a
comparison with the stellar evolution models since nearly every study used
a different distance modulus and reddening.

We now consider each of the isochrone fits listed in Table 3.  While
theoretical stellar evolution models differ in numerous ways, including
the convective mixing length used in the near-surface regions, whether or
not diffusion is included, the detailed translation from the theoretical
temperature - luminosity plane to the observational color - magnitude
plane, and in the prescription for convective core overshoot if used,
there remains a clear difference between those studies that do and do not
incorporate core convective overshoot.

For historical comparison, and since so many recent papers adopt some of
the parameters derived by Anthony-Twarog \etal (1990), we report their
results here, even though their cluster parameters (but not their
photometry) are now superseded by other studies.  As remarked above,
Anthony-Twarog \etal found a problematic fit with VandenBerg's (1985)
isochrones, from which they derived an age of $3.4 \pm 0.6$ Gyrs and
(m$-$M)$_{\rm V} = 11.95$.  They assumed [Fe/H] $= -0.4$ and E(B$-$V) $=
0.05$, both of which are reasonable values according to nearly all
subsequent efforts.  Anthony-Twarog \etal argued that the poor fit between
their photometry and VandenBerg's models likely indicated the need for
models including core convective overshoot.  Given the development in
stellar evolution modeling and input physics in the last fifteen years,
particularly updated opacity tables, the application of VandenBerg (1985)
models in the Anthony-Twarog \etal study serves more as a starting point
for the issue of convective core overshoot than as a definitive
statement.  This isochrone age is also high by all modern estimates, as
our discussion will reveal.

Castellani, Chieffi, \& Straniero (1992) use canonical stellar models and
derive a cluster age of $1.7$ Gyrs, with E(B$-$V) = $0.02$ and
(m$-$M)$_{\rm V} = 11.95$.  Unfortunately, these results cannot be
directly compared to those of other groups since their study was meant as
a test of theory, and they only compared the NGC 2420 photometry to solar
metallicity isochrones.  Nonetheless, their derived age is consistent with
the most recent ages derived by other groups using canonical models.

Dominguez \etal (1999) concluded that the complex shape of the main
sequence turn-off region in NGC 2420 is not due to convective core
overshoot, as argued by many others, but rather to the confusing
photometric locations of multiple stars.\footnote{The effect of multiple
stars on the turn-off region can be independently verified by radial
velocity techniques (e.g. Daniel \etal 1994), but this test has not yet
been performed for NGC 2420.}  With this interpretation they derived an
age of $1.6 \pm 0.2$ Gyrs and (m$-$M)$_{\rm V} = 12.19$.  Dominguez \etal
also noted the consistency of their isochrone age with the age implied by
the coolest WDs we reported in paper 1.

Castellani, degl'Innocenti, \& Marconi (1999), in a study of mixing length
theory (surface convection) and models incorporating diffusion but not
incorporating core convective overshoot, derive an age for NGC 2420 of
$1.5 \pm 0.1$ Gyr, along with E(B$-$V) $\approx 0.16$, and (m$-$M)$_{\rm
V} = 12.4$.  In this case, the high reddening and distance modulus make
their age result suspect.

Carraro \& Chiosi (1994) applied their convective core overshoot models to
NGC 2420 and derived an age of $2.1$ Gyrs, E(B$-$V) = $0.08$, and
(m$-$M)$_{\rm V} = 11.80$.  For the reasons discussed above, their
distance modulus may be too low, and their reddening value is likely to be
modestly too high.

Demarque \etal (1994) fit the photometry of Anthony-Twarog \etal with
their updated models with and without convective core overshoot.  They
concluded that the photometry required models with core overshoot, with an
overshoot parameter P$_{mix} = 0.23$ H$_p$ and age = $2.4 \pm 0.2$ Gyrs.
Their distance and reddening determinations, as discussed above, appear to
be of high quality.

In a series of papers Pols and collaborators (Schroder, Pols, \& Eggleton,
1997; Pols \etal 1997; Pols \etal 1998) tested their canonical and
overshoot models against giants of known mass, double-lined spectroscopic
binaries, and open clusters.  They recommend the use of convective core
overshoot models for solar metallicity stars with main sequence masses
$\geq 1.5$ solar masses.  This limit of $\approx 1.5$ solar masses scales
inversely with metallicity and they note that for the metallicity of NGC
2420 the limit is probably slightly less than $1.4$ solar masses.  In
their fit to NGC 2420 Pols \etal decisively favor the overshoot models and
find an age of $2.35$ Gyrs and a turn-off mass of $1.47$ solar masses,
after adopting [Fe/H] = $-0.42$, E(B$-$V) = $0.05$, and (m$-$M)$_{\rm V} =
11.95$.

Twarog \etal (1999) find an age of $1.9 \pm 0.2$ or $2.2 \pm 0.2$ Gyrs
based on either the models of Bertelli \etal (1994) or Schaerer \etal
(1993) and Schaller \etal (1992) respectively, all of which employ
convective core overshoot.  While the real goal of the Twarog \etal study
was not to determine the isochrone age of NGC 2420, but rather to study
the absolute magnitude of the red giant branch clump, their careful study
appears to yield good isochrone fits, and good values for the cluster
distance and reddening.  While the Bertelli \etal isochrone age is the
youngest overshoot age derived for NGC 2420, it is still older than any of
the modern canonical isochrone fits.

In Figure 8 we present a summary of the modern stellar evolution isochrone
ages for NGC 2420 along with our WD cooling age as a function of the
assumed distance modulus.  The solid line is our best fit WD cooling age.
The two dashed lines represent the $\pm 1 \sigma$ differences from our
best age of $0.2$ Gyrs.  The canonical isochrone ages are indicated by the
open squares, whereas the convective overshoot isochrone ages are
indicated by the filled circles.  We use the age errors given by the
authors when given (see Table 3), and otherwise assume an age uncertainty
of $\pm 0.2$ Gyrs.  Most of the overshoot model ages agree with our WD
cooling ages.  The canonical models, on the other hand, produce ages that
are generally in conflict with the WD cooling ages.  The oldest canonical
model clearly is inconsistent with the WD cooling ages, but this is the
now-outdated model of VandenBerg (1985), and therefore the poor agreement
is not surprising.  The canonical model at (m$-$M)$_{\rm V} = 11.95$ and
$1.7$ Gyrs is from the Castellani \etal (1992) study employing solar
metallicity isochrones, so the disagreement here is not surprising
either.  Of the remaining two canonical studies (Castellani \etal 1999 and
Dominguez \etal 1999), both produce isochrone ages consistent with the WD
cooling ages, though both employ high distance moduli.  The distance
discrepancy for the Dominguez \etal study may be minor, however.

As a final note on the comparison between isochrone and WD cooling ages,
an incorrect distance modulus can inappropriately alter the assumed
physics involved in stellar evolution, since the assumed stellar
luminosity and thereby the assumed stellar mass depends on the assumed
cluster distance.  Therefore, the last word on the question of canonical
versus core overshoot models for this cluster may have to await a precise
and agreed-upon determination of the cluster distance modulus.  In turn, a
precise distance modulus is likely to require an improved cluster
metallicity value.  In the meantime, since both WD cooling ages and
isochrone ages scale similarly with the assumed distance, we have applied
the fairest comparison we can.  Once a precise cluster distance modulus is
determined, our WD photometry provides a strict consistency check of any
isochrone fitting.

\subsection{The Limit of the Main Sequence}

While our observations were designed to measure the luminosities of the
faintest cluster WDs, they also revealed a CMD that probes the cluster
main sequence from V = $16$ to $26$.  This apparent magnitude range
corresponds to the mass range of $1.2$ to $0.15$ solar masses, based on
the empirical mass - luminosity calibration of Henry \& McCarthy (1993).
We do not derive a cluster luminosity function or initial mass function
from these data, as done by von Hippel \etal (1996) from the cycle 4 data,
as the increase in numbers of stars does not warrant a re-examination of
this subject.  However, the deeper exposures and the larger field of view
allowed us to detect candidate cluster main sequence stars significantly
fainter than those studied by von Hippel \etal (1996).  Because of the
current interest in comparing the photometric properties of faint main
sequence stars in clusters at different metallicity, we tabulate the main
sequence ridge line in Table 4.  The ridge line was determined by fitting
a fourth-order polynomial to the photometry for those stars that appear to
lie along the well-represented portion of the single-star main sequence,
between M$_{\rm V} \approx 9$ to $14$ (V $\approx 21$ to $26$, mass
$\approx 0.58$ to $0.15$ solar masses), assuming (m$-$M)$_{\rm V} = 12.10$
and E(B$-$V) = $0.04$.  The quality of the fit can be seen by examining
Figure 7a, where the ridge line is represented by ``+'' symbols.  We
remind the reader that all of these stars are redder than the red limit of
the Holtzman \etal (1995) photometric transformations, V$-$I $< 1.5$.

We can make tentative comments about cluster members of even lower mass
than $0.15$ solar masses.  At the faint limit of our photometry we found
three objects that passed our I-band SExtractor stellarity index cut at
$0.78$ (with values of $0.87$, $0.95$, and $0.96$), but which were too
faint in the V-band to provide a reliable centroid, and thus were rejected
as poor matches between the V- and I-band photometry lists.  These objects
were measured to have I $\approx 23.8$, $24.6$, and $24.7$.  These
magnitudes are approximate due to their low signal-to-noise and uncertain
color correction.  The V-band photometry is even more uncertain.  Table 5
lists the I- and V-band magnitudes for these three objects, along with
their Right Ascension and Declination (J2000.0).  The brightest of these
three objects is likely to be a cluster member, based on its position in
the CMD.  It is, in fact, the same color as and only $0.2$ mag fainter
than the two reddest main sequence candidates plotted in our CMDs.  The
mass of this object, assuming it is a cluster member, is $\approx 0.12$
solar masses, based on the calibration of Henry \& McCarthy (1993).  The
two fainter objects, if cluster main sequence stars, would have slightly
lower mass, $\approx 0.11$ solar masses.  Since the completeness level at
these faint magnitudes is only a few percent, it appears likely that NGC
2420 contains many very faint stars near the Hydrogen-burning limit.

\section{Conclusion}

We have used deep HST WFPC2 observations of two fields in NGC 2420 to
produce a cluster CMD down to V $\approx 27$.  After imposing
morphological selection criteria we find eight candidate white dwarfs in
NGC 2420.  Our completeness estimates indicate that we have found the
terminus of the WD cooling sequence.  We argue that the cluster distance
modulus is likely to be close to $12.10$ with E(B$-$V) = $0.04$.  With
these parameters we find a white dwarf cooling age for NGC 2420 of $2.0
\pm 0.20$ ($1 \sigma$) Gyrs.  The $0.20$ Gyr uncertainty includes errors
in the photometry, sequence fitting, precursor time scales, and
theoretical WD cooling time scales.

We derive cluster WD ages for a variety of distances to directly compare
the WD age with the many main sequence evolution ages for NGC 2420.  We
find that most of the stellar evolution models that incorporate convective
overshoot derive ages which agree with our WD cooling ages.  The canonical
models, on the other hand, largely produce ages that are in conflict with
the WD cooling ages.  An exception to this tendency is the canonical
isochrone fit of Dominguez \etal (1999), which results in an age that is
consistent with the WD cooling age, but with a distance modulus that may
be too high.  The final word on the question of canonical versus core
overshoot models for this cluster may have to await a precise and
agreed-upon determination of the cluster distance modulus, which in turn,
will likely require an improved cluster metallicity value.

\acknowledgments

It is a pleasure to thank Ken Mighell for the use of the CCDCAP photometry
package and for many useful consultations, and Andy Fruchter for guidance
with the drizzle package.  We thank Leandro Althaus, Brad Hansen, Jarrod
Hurley, and Matt Wood for providing their computer-readable model
calculations or code and for helpful guidance.  We also thank Pierre
Demarque for helpful guidance.

Support for this work was provided by NASA through grant number GO-6424
from the Space Telescope Science Institute, which is operated by the
Association of Universities for Research in Astronomy, Inc., under NASA
contract NAS5-26555.

This research has made extensive use of NASA's Astrophysics Data System
Abstract Service.

\eject

\newpage

\figcaption[fig1.ps]
{The NGC 2420 CMD.  All objects with $\sigma_{\rm V}$ and $\sigma_{\rm I}
\leq 0.15$ mags are plotted, along with their $1 \sigma$ error bars.  No
cut is made on image morphology.}

\figcaption[fig2.ps]
{Apparent V mag versus the SExtractor (Bertin \& Arnouts 1996)
``stellarity index''.  The stellarity index ranges from $1.0$ for definite
unresolved objects to $0.0$ for definite resolved objects.  The horizontal
dashed line at $0.78$ is our stellar classification threshold.  The
vertical arrows on the left indicate ranges for resolved and unresolved
objects, as well as image defects and noise.  {\it All} objects with a
stellarity index $\geq 0.9$ are clearly unresolved and {\it all} objects
with a stellarity index $\leq 0.2$ are clearly resolved.  These areas of
certainty are indicated by the solid vertical arrows.  The dashed vertical
arrows indicate areas of parameter space where {\it some} objects are
either clearly resolved or clearly unresolved.}

\figcaption[fig3.ps]
{The NGC 2420 CMD after removal of all resolved objects and image
defects.  Only objects with $\sigma_{\rm V}$ and $\sigma_{\rm I} \leq
0.15$ mags and with a stellarity index $\geq 0.78$ are plotted.  The line
is a cooling track for a $0.7$ solar mass C-O WD (Benvenuto \& Althaus
1999) placed at a cluster distance modulus of $12.10$.}

\figcaption[fig4.ps]
{The CMD for the Galaxy model of Reid \& Majewski (1993) for the location
({\it l} = $198$, {\it b} = $20$), sky coverage, and reddening of our NGC
2420 fields.}

\figcaption[fig5.ps]
{The WD portion of the CMD displaying all possible WD candidates.  Objects
were selected based on their photometric proximity ($1.5 \sigma$) to the
model cooling tracks for $0.6$ to $1.0$ solar mass C-O WDs (Benvenuto \&
Althaus 1999) placed at the cluster distance.  The symbols indicate the
stellarity index range. Objects with stellarity index $\leq 0.2$ are
obviously resolved to the eye and therefore are not plotted.  The dotted
lines in the lower left are the probabilities of finding objects as a
function of luminosity, with the first and second numbers indicating
completeness in fields 1 and 2, respectively.  Each object is labeled with
the field in which it was found.}

\figcaption[fig6.ps]
{
The luminosity functions of objects from Figure 5 and field 2 above the
stellarity index thresholds indicated in each panel.  The cross-hatched
histogram presents the observed LFs, whereas the unfilled histogram
presents the completeness-corrected LFs.  The final, corrected LF bin of
each of panels c through f contains $200$ objects.  Overplotted on each
panel is the WD LF of Richer \etal (1998) for the $\approx 4$ Gyr open
cluster M67, along with error bars due to both counting statistics and
background subtraction errors.
}

\figcaption[fig7.ps]
{a. The NGC 2420 CMD with our preferred distance modulus and reddening
removed.  The $0.5, 1.0, 1.5, 2.0, 2.5,$ and $3.0$ Gyrs isochrones (Ahrens
1999) and $0.6$ solar mass WD cooling track (Wood 1992) are over-plotted.
The main sequence ridge line is indicated by ``+'' symbols.  b. Same as a,
but only showing the WD region of the CMD.}

\figcaption[fig8.ps]
{A comparison between all modern isochrone ages determined for NGC 2420
and our WD cooling age as a function of assumed distance modulus.  The
solid line is our best fit WD cooling age.  The two dashed lines represent
the $\pm 1 \sigma$ uncertainties of $0.2$ Gyrs.  The canonical isochrone
ages are plotted as open squares and the convective overshoot isochrone
ages are plotted as filled circles.}

\clearpage

%Table 1
\begin{deluxetable}{lccccc}
\tablewidth{0pt}
\tablecaption{Log of Observations}
\tablehead{
\colhead{Date} & \colhead{Exposures} & \colhead{Filter} & \colhead{Sky} &
\colhead{Field} & \colhead{Cycle}  \\
\phantom{1234}(1) & (2) & (3) & (4) & (5) & (6)}
\startdata
May 18, 1994 & 4 $\times$ 900     & F555W & 33.6 & 1 & 4 \\
May 19, 1994 & 4 $\times$ 900     & F814W & 33.2 & 1 & 4 \\
Apr  2, 1996 & 2 $\times$ 1200    & F555W & 16.8 & 1 & 6 \\
Apr  2, 1996 & 2 $\times$ 1200    & F555W & 17.0 & 1 & 6 \\
Apr  3, 1997 & 3 $\times$ 700/800 & F555W & 11.7 & 2 & 6 \\
Apr  3, 1997 & 3 $\times$ 700/800 & F555W & 11.7 & 2 & 6 \\
Apr  3, 1997 & 3 $\times$ 700/800 & F555W & 11.6 & 2 & 6 \\
Apr  3, 1997 & 3 $\times$ 700/800 & F814W & 11.1 & 2 & 6 \\
Apr  3, 1997 & 3 $\times$ 700/800 & F814W & 11.1 & 2 & 6
\enddata
\end{deluxetable}

%Table 2
%
\begin{deluxetable}{lccrcll}
\tablewidth{0pt}
\tablecaption{White Dwarfs}
\tablehead{
\colhead{ID} & \colhead{V} & \colhead{$\sigma_{\rm V}$} & 
\colhead{V$-$I} & \colhead{$\sigma_{\rm V-I}$} & \colhead{RA} & \colhead{Dec} \\
\phantom{1}(1) & (2) & (3) & (4)\phantom{1} & (5) & \phantom{123}(6) & \phantom{123}(7)}
\startdata
WD1 & 21.43 & 0.003 & -0.23 & 0.009 & 7 38 10.62 & 21 32 33.5 \\
WD2 & 22.52 & 0.005 & -0.16 & 0.016 & 7 38 08.07 & 21 33 08.8 \\
WD3 & 23.66 & 0.010 & -0.08 & 0.025 & 7 38 26.54 & 21 35 13.1 \\
WD4 & 24.15 & 0.015 &  0.08 & 0.044 & 7 38 10.85 & 21 31 56.6 \\
WD5 & 25.45 & 0.032 &  0.39 & 0.069 & 7 38 31.26 & 21 35 56.6 \\
WD6 & 25.64 & 0.038 &  0.75 & 0.067 & 7 38 29.68 & 21 36 45.4 \\
WD7 & 25.68 & 0.038 &  0.46 & 0.079 & 7 38 30.50 & 21 34 31.3 \\
WD8 & 25.69 & 0.039 &  0.63 & 0.073 & 7 38 30.12 & 21 36 34.0
\enddata
\end{deluxetable}

%Table 3
%
\begin{deluxetable}{lccccccc}
\tablewidth{0pt}
\tablecaption{Ages Derived from Isochrone Fits}
\tablehead{
\colhead{Age(Iso)} & \colhead{Overshoot?} & \colhead{(m$-$M)$_{\rm V}$} & 
\colhead{E(B$-$V)} & \colhead{Derived?} & \colhead{[Fe/H]} & 
\colhead{Ref} & \colhead{Age(WD)} \\
\phantom{12}(1) & (2) & (3) & (4) & (5) & (6) & (7) & (8)}
\startdata
3.4 (0.6) & N  & 11.95 & 0.05             & D  & -0.40           & AT90 & 2.3 \\
1.7       & N  & 11.95 & 0.02             & D  & \phantom{.}0.00 & C92  & 2.3 \\
1.6 (0.2) & N  & 12.19 & 0.05             & D  & -0.40           & D99  & 1.9 \\
1.5 (0.1) & N  & 12.40 & 0.16             & D  & -0.40           & C99  & 1.6 \\
2.1       & Y  & 11.80 & 0.08             & D  & -0.42           & CC94 & 2.5 \\
2.4 (0.2) & Y  & 12.05 & \phantom{1}0.045 & D  & -0.30           & D94  & 2.1 \\
2.35      & Y  & 11.95 & 0.05             & A  & -0.42           & P98  & 2.3 \\
1.9 (0.2) & Y  & 12.15 & 0.04             & D  & -0.29           & T99a & 2.0 \\
2.2 (0.2) & Y  & 12.15 & 0.04             & D  & -0.29           & T99b & 2.0
\enddata
\tablerefs{
AT90 = Anthony-Twarog \etal 1990,
C92 = Castellani \etal 1992,
C99 = Castellani \etal 1999,
CC94 = Carraro \& Chiosi 1994,
D94 = Demarque \etal 1994,
D99 = Dominguez \etal 1999,
P98 = Pols \etal 1998,
T99a = Twarog \etal 1999 employing Bertelli \etal 1994 isochrones, and
T99b = Twarog \etal 1999 employing Schaller \etal 1992 \& Schaerer \etal
     1993 isochrones.
}
\end{deluxetable}

%Table 4
%
\begin{deluxetable}{cr}
\tablewidth{0pt}
\tablecaption{Main Sequence Ridge Line}
\tablehead{
\colhead{V$-$I} & \colhead{\phantom{1}M$_{\rm V}$} \\
(1) & (2)\phantom{1}}
\startdata
1.8 &  8.66 \\
1.9 &  9.06 \\
2.0 &  9.49 \\
2.1 &  9.95 \\
2.2 & 10.42 \\
2.3 & 10.89 \\
2.4 & 11.36 \\
2.5 & 11.82 \\
2.6 & 12.25 \\
2.7 & 12.66 \\
2.8 & 13.02 \\
2.9 & 13.33 \\
3.0 & 13.59
\enddata
\end{deluxetable}

%Table 5
%
\begin{deluxetable}{lccll}
\tablewidth{0pt}
\tablecaption{Faint Red Stars}
\tablehead{
\colhead{ID} & \colhead{I} & \colhead{V} & \colhead{RA} & \colhead{Dec} \\
\phantom{1}(1) & (2) & (3) & \phantom{123}(4) & \phantom{123}(5)}
\startdata
MS1 & 23.8 & 26.9    & 7 38 30.94 & 21 34 44.9 \\
MS2 & 24.6 & 27.5    & 7 38 31.62 & 21 36 37.3 \\
MS3 & 24.7 & \nodata & 7 38 23.74 & 21 35 30.4
\enddata
\end{deluxetable}


\begin{references}

\reference{}
Ahrens, T.J. 1999, Master's Thesis, Fl. Inst. of Tech.

\reference{}
Anderson, J., \& King, I.R. 1999, \pasp, 111, 1095

\reference{}
Anthony-Twarog, B.J., Kaluzny, J., Shara, M.M., \& Twarog, B.A. 1990, \aj,
99, 1504

\reference{}
Benvenuto, O.G., \& Althaus, L.G. 1999, \mnras, 303, 30

\reference{}
Bertelli, G., Bressan, A., Chiosi, C., Fagotto, F., \& Nasi, E. 1994,
\aaps, 106, 275

\reference{}
Bertin, E., \& Arnouts, S. 1996, \aaps, 117, 393

\reference{}
Canuto, V.M. 1999, \apj, 524, 311

\reference{}
Canuto, V.M., \& Dubovikov, M. 1998, \apj, 493, 834

\reference{}
Cardelli, J.A., Clayton, G.C., \& Mathis, J.S. 1989, \apj, 345, 245

\reference{}
Carraro, G., \& Chiosi, C. 1994, \aap, 287, 761

\reference{}
Casertano, S., \& Mutchler, M. 1998, WFPC2 Instrument Science Report,
98-02

\reference{}
Castellani, V., Chieffi, A., \& Straniero, O. 1992, \apjs, 78, 517

\reference{}
Castellani, V., degl'Innocenti, S., \& Marconi, M. 1999, \mnras, 303, 265

\reference{}
Claver, C.F. 1995, PhD Thesis, The University of Texas at Austin

\reference{}
Daniel, S.A., Latham, D.W., Mathieu, R.D., \& Twarog, B.A. 1994, \pasp,
106, 281

\reference{}
Demarque, P., Sarajedini, A., \& Guo, X.-J. 1994, \apj, 426, 165

\reference{}
Dominguez, I., Chieffi, A., Limongi, M., \& Straniero, O. 1999, \apj, 524,
226

\reference{}
Friel, E.D., \& Janes, K.A. 1993, \aap, 267, 75

\reference{}
Fruchter, A.S., \& Hook, R.N. 1998, xxx.lanl.gov/abs/astro-ph/9808087

\reference{}
Gonzaga, S., Ritchie, C., Baggett, S., Whitmore, B. \& Mutchler, M. 1999,
Standard Star Monitoring Memo \#3

\reference{}
Hansen, B.M.S. 1999, \apj, 520, 680

\reference{}
Henry, T.J., \& McCarthy, D.W. 1993, \aj, 106, 773

\reference{} 
Holtzman, J.A., Burrows, C.J., Casertano, S., Hester, J.J., Watson,
A.M., \& Worthy, G.S. 1995, \pasp, 107, 1065

\reference{} 
Hurley, J.R., Pols, O.R., \& Tout, C.A. 2000, \mnras, in press

\reference{} 
Iben, I., \& Tutukov, A.V. 1984, \apj, 282, 615

\reference{} 
Koester, D., \& Reimers, D. 1985, \aap, 153, 260

\reference{} 
Koester, D., \& Reimers, D. 1996, \aap, 313, 810

\reference{} 
Krist, J. 1995, in Astronomical Data Analysis Software and Systems IV,
eds. R.A. Shaw, H.E. Payne, \& J.J.E. Hayes, ASP Conference Series, (San
Francisco: ASP), 77, 349

\reference{}
Leggett, S.K., Ruiz, M.T., \& Bergeron, P. 1998, \apj, 497, 294

\reference{}
Liebert, J., Dahn, C.C., \& Monet, D.G. 1988, \apj, 332, 891

\reference{}
Mighell, K.J. 1997, \aj, 114, 1458

\reference{}
Oswalt, T.D., Smith, J.A., Wood, M.A., \& Hintzen, P. 1996, \nat, 382, 692

\reference{}
Pinsonneault, M.H., Stauffer, J., Soderblom, D.R., King, J.R., \& Hanson,
R.B. 1998, \apj, 504, 170

\reference{}
Pols, O.R., Tout, C.A., Schroder, K.-P., Eggleton, P.P., \& Manners, J.
1997, \mnras, 289, 869

\reference{}
Pols, O.R., Schroder, K.-P., Hurley, J.R., Tout, C.A., \& Eggleton, P.P.
1998, \mnras, 298, 525

\reference{}
Schmidt, M. 1959, \apj, 129, 243

\reference{}
Reid, I.N., \& Majewski, S.R. 1993, \apj, 409, 635

\reference{}
Reid, I.N., Yan, L., Majewski, S., Thompson, I., \& Smail, I. 1996, \aj,
112, 1472

\reference{}
Richer, H.B., \etal 1997, \apj, 484, 741

\reference{}
Richer, H.B., Fahlman, G.G., Rosvick, J., \& Ibata, R. 1998, \apj, 504,
L91

\reference{}
Salaris, M., Dominguez, I., Garcia-Berro, E., Hernanz, M., Isern, J., \&
Mochkovitch, R. 1997, \apj, 486, 413

\reference{}
Schaerer, D., Meynet, G., Maeder, A., \& Schaller, G. 1993, \aaps, 98, 523

\reference{}
Schaller, G., Schaerer, D., Meynet, G., \& Maeder, A. 1992, \aaps, 96, 269

\reference{}
Schroder, K.-P., Pols, O.R., \& Eggleton, P.P. 1997, \mnras, 285, 696

\reference{}
Stetson, P.B. 1998, \pasp, 110, 1448

\reference{}
Twarog, B.A., Anthony-Twarog, B.J., \& Bricker, A.R. 1999, \aj, 117, 1816

\reference{}
van Altena, W.F., \& Jones, B.F. 1970, \aap, 8, 112

\reference{}
VandenBerg, D.A. 1985, \apjs, 58, 532

\reference{}
von Hippel, T. 1998, \aj, 115, 1536

\reference{}
von Hippel, T., Gilmore, G., \& Jones, D.H.P. 1995, \mnras, 273, L39
(paper 1)

\reference{}
von Hippel, T., Gilmore, G., Tanvir, N., Robinson, D., \& Jones, D.H.P.
1996, \aj, 112, 192

\reference{}
von Hippel, T., \& Sarajedini, A. 1998, \aj, 116, 1789

\reference{}
Whitmore, B., Heyer, I., \& Casertano, S. 1999, \pasp, 111, 1559 

\reference{}
Winget, D.E., Hansen, C.J., Liebert, J., van Horn, H.M., Fontaine, G.,
   Nather, R.E., Kepler, S.O., \& Lamb, D.Q. 1987, \apj, 315, L77

\reference{}
Wood, M.A. 1992, \apj, 386, 539

\end{references}
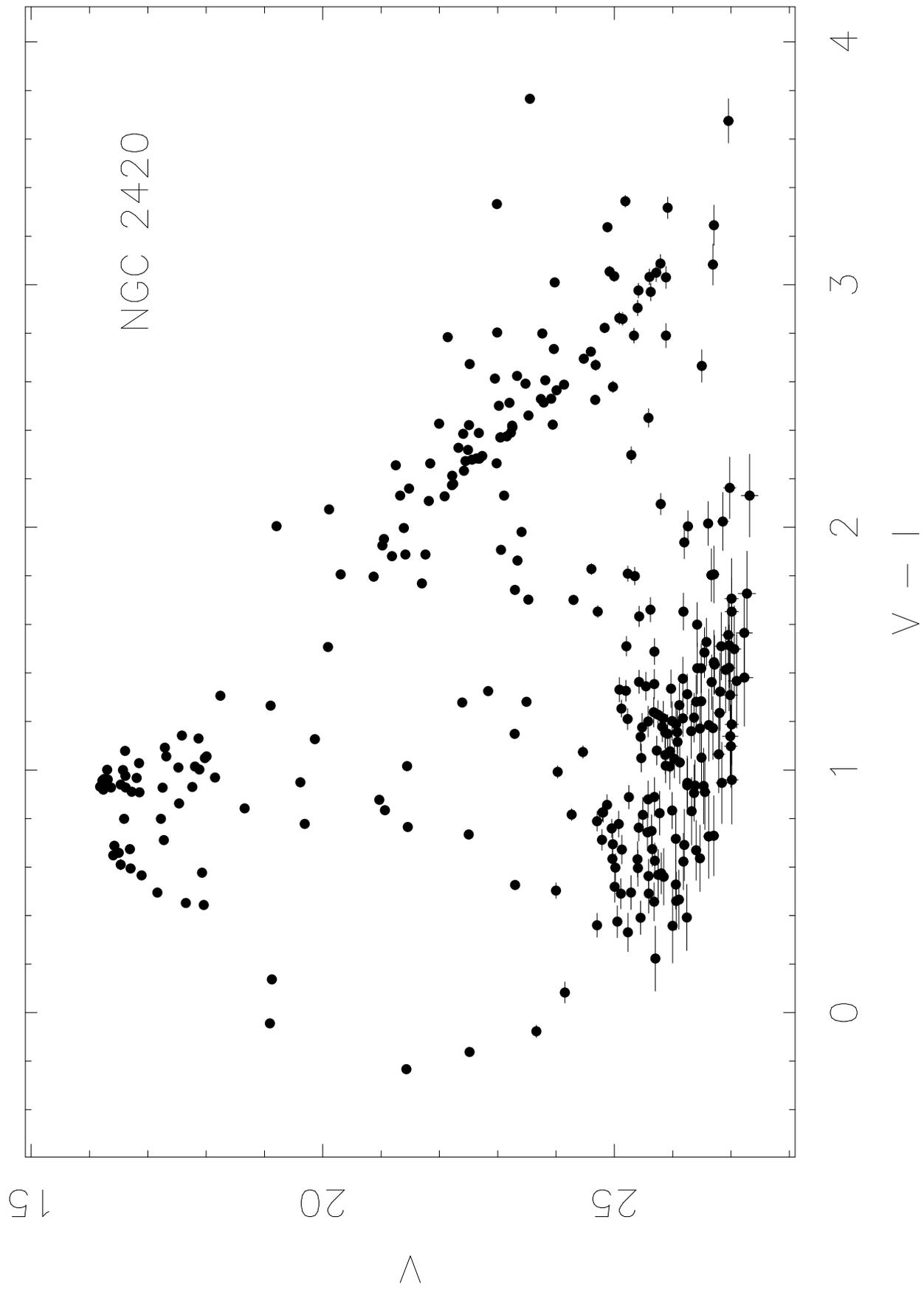
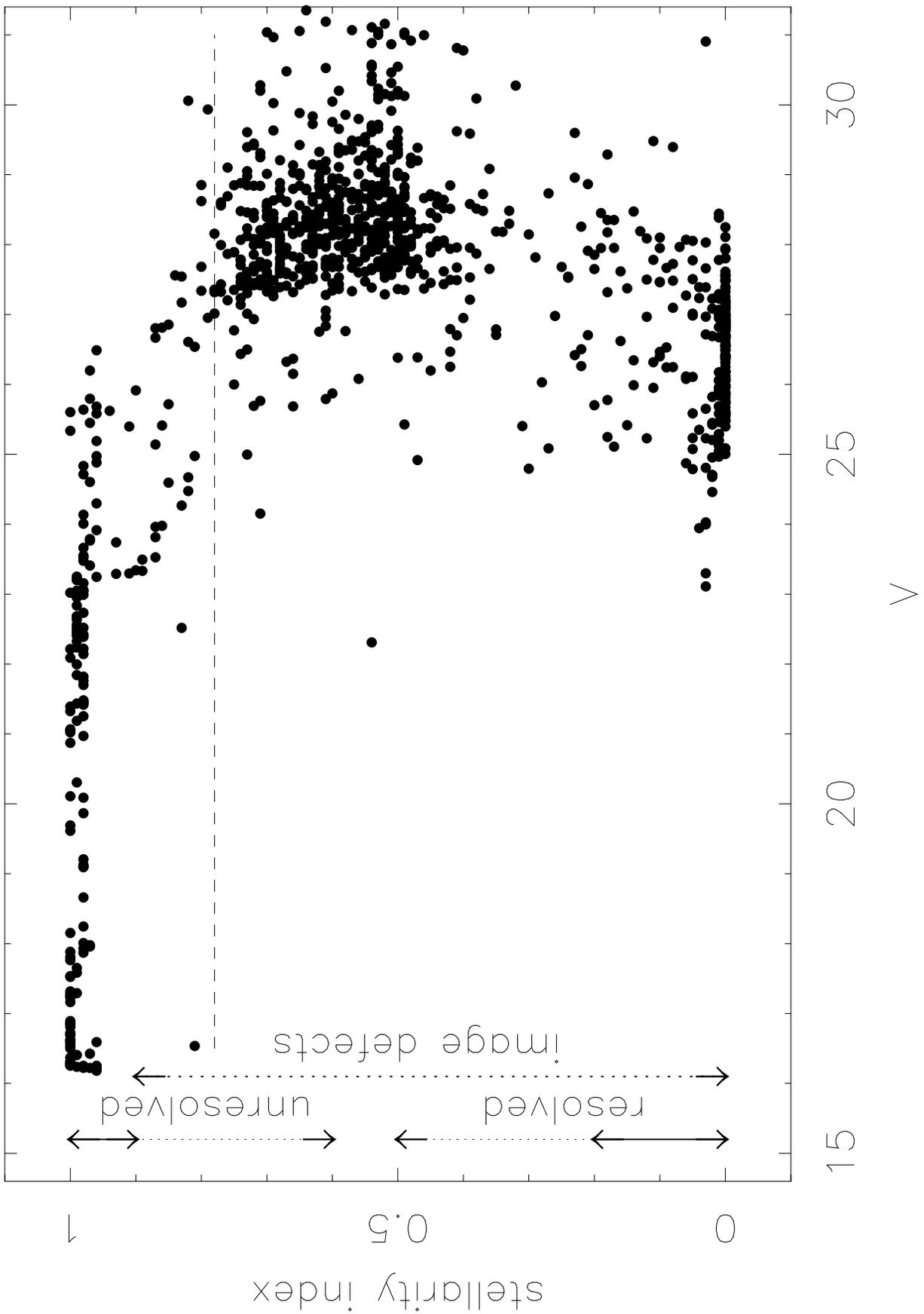
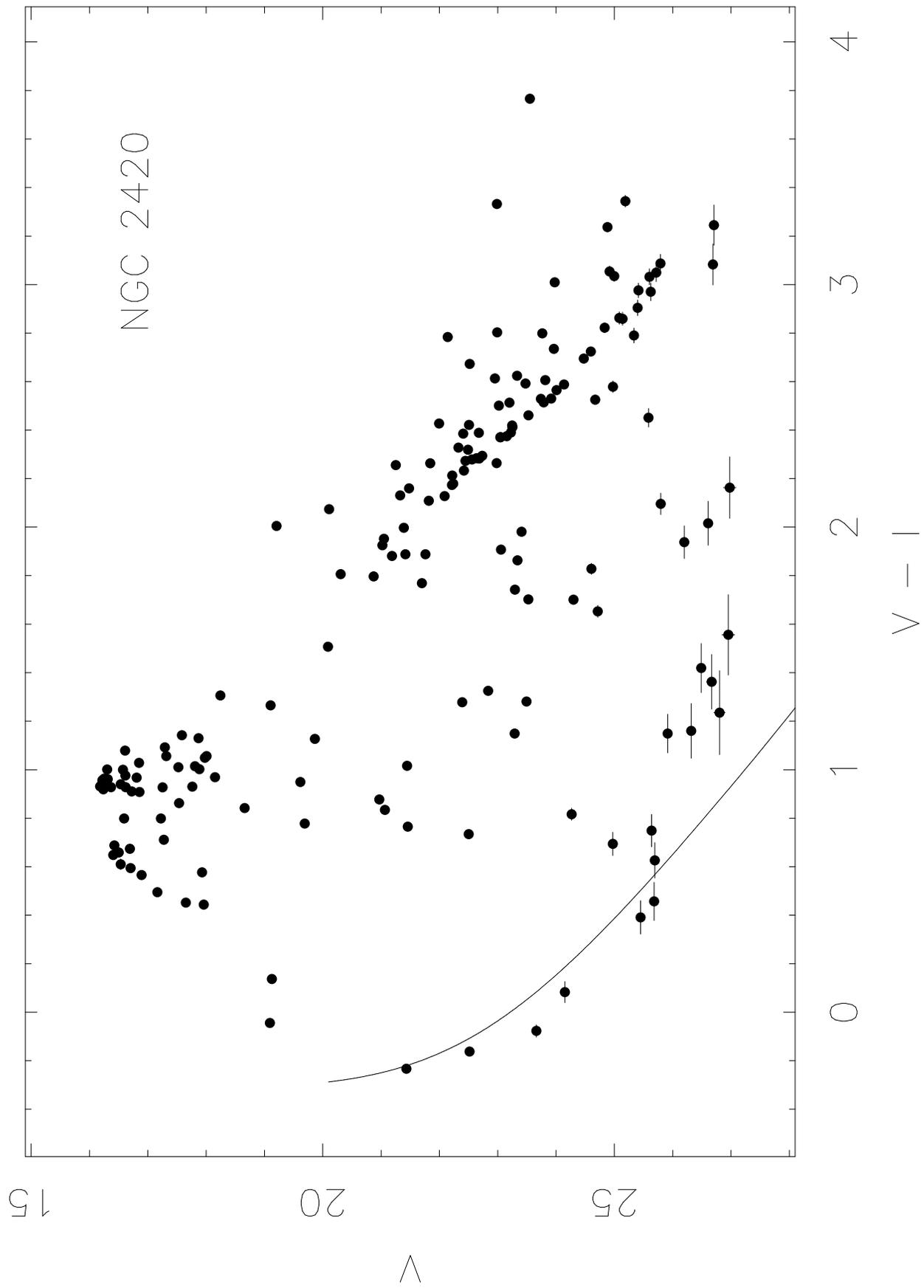
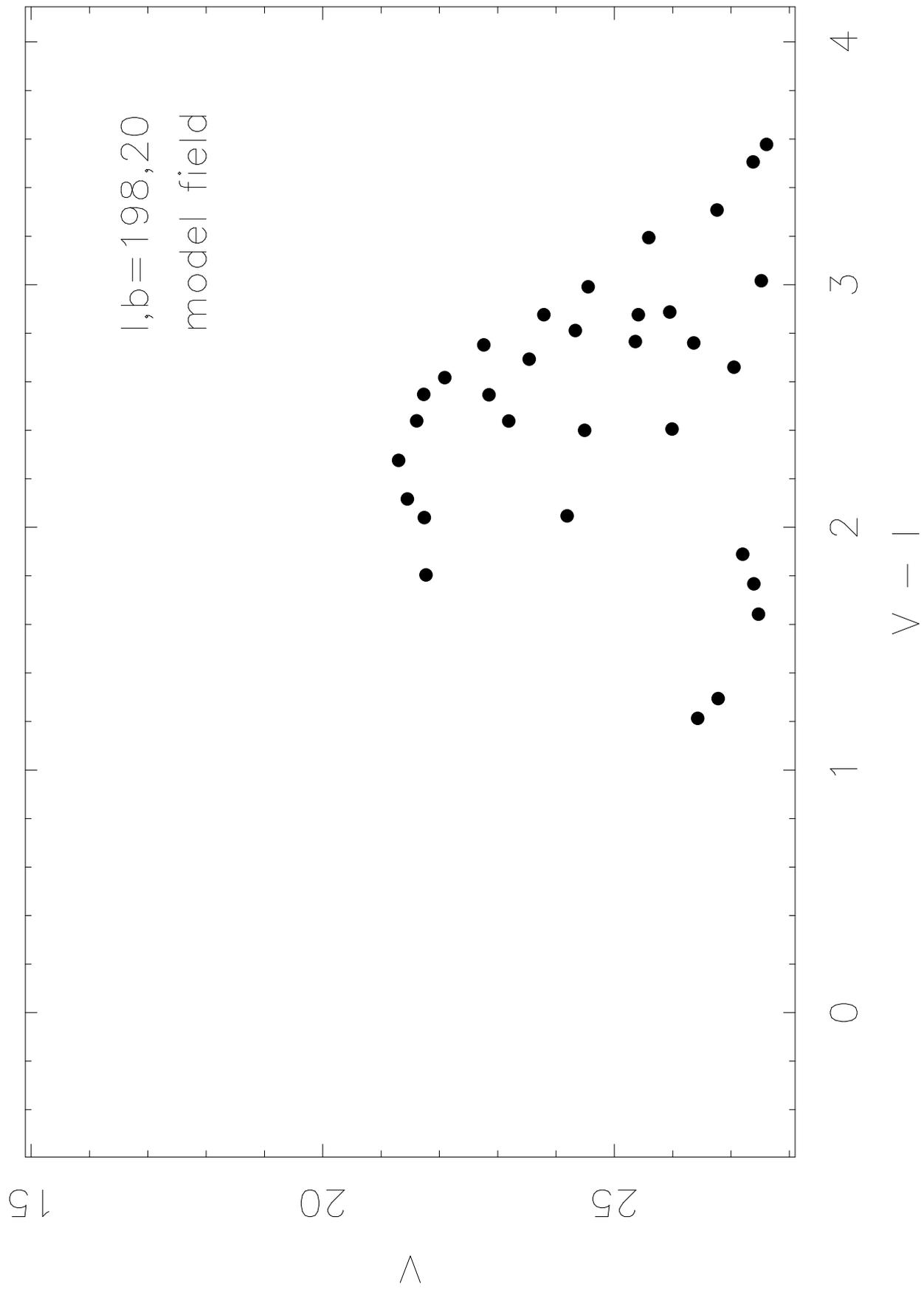
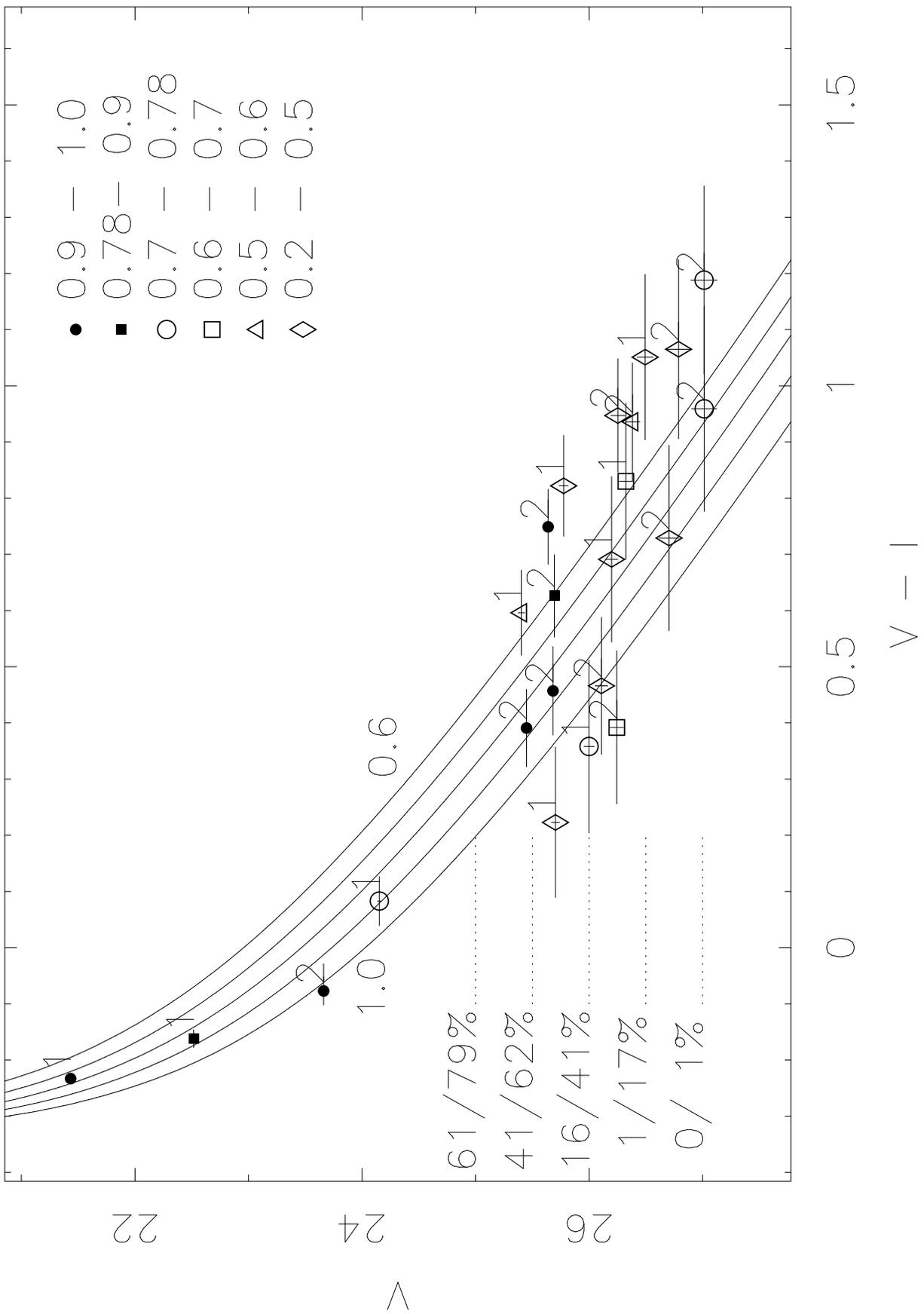
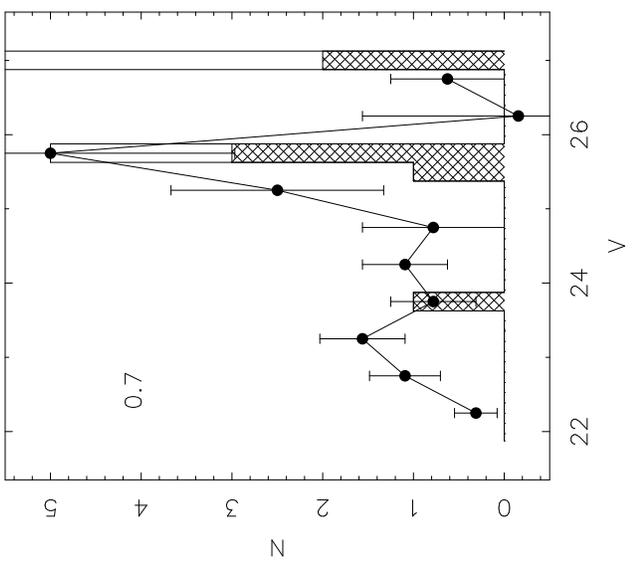
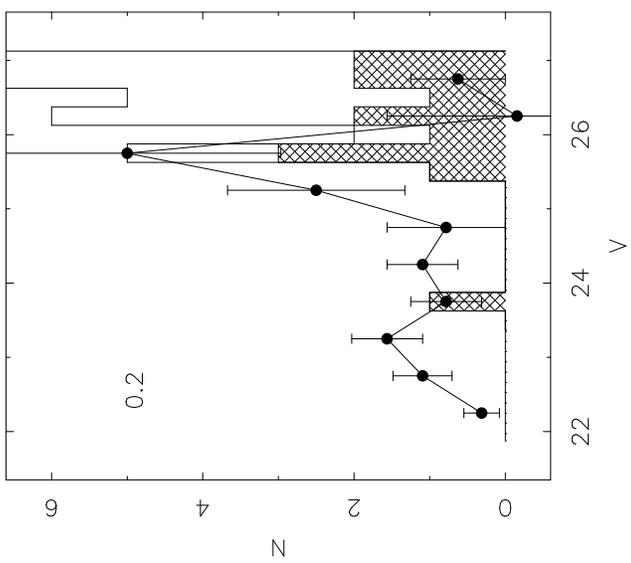
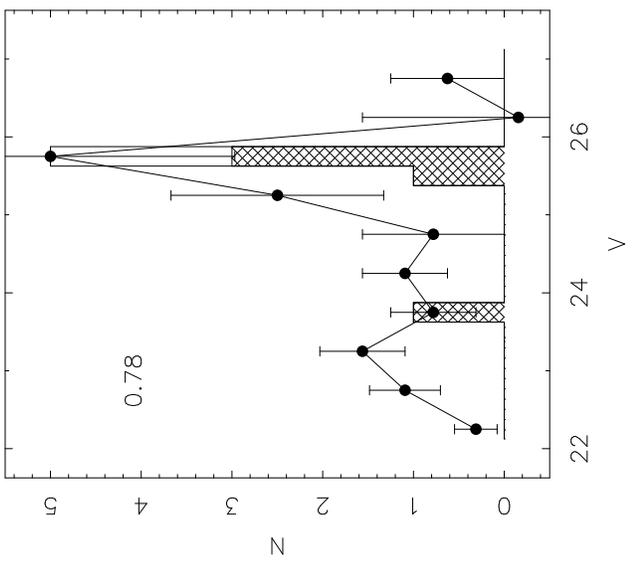
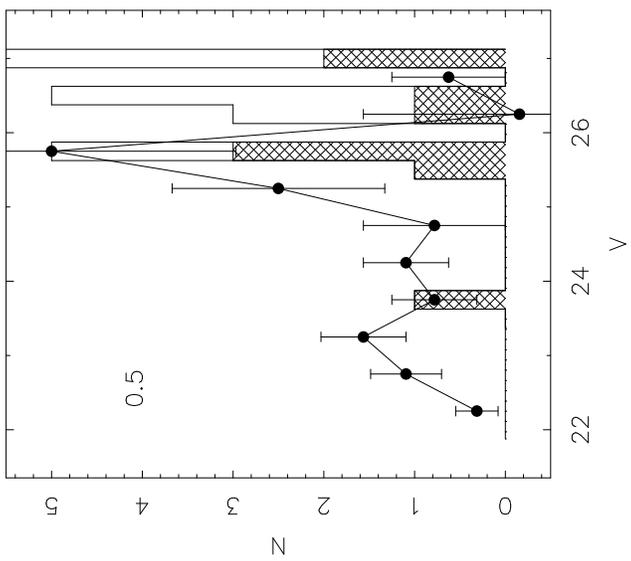
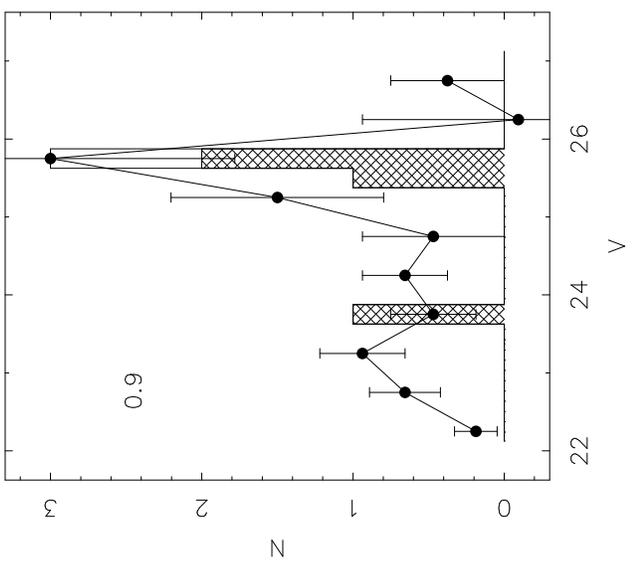
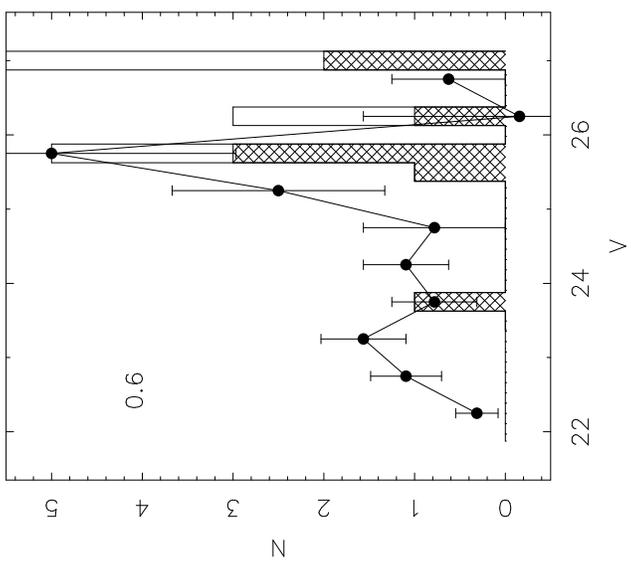
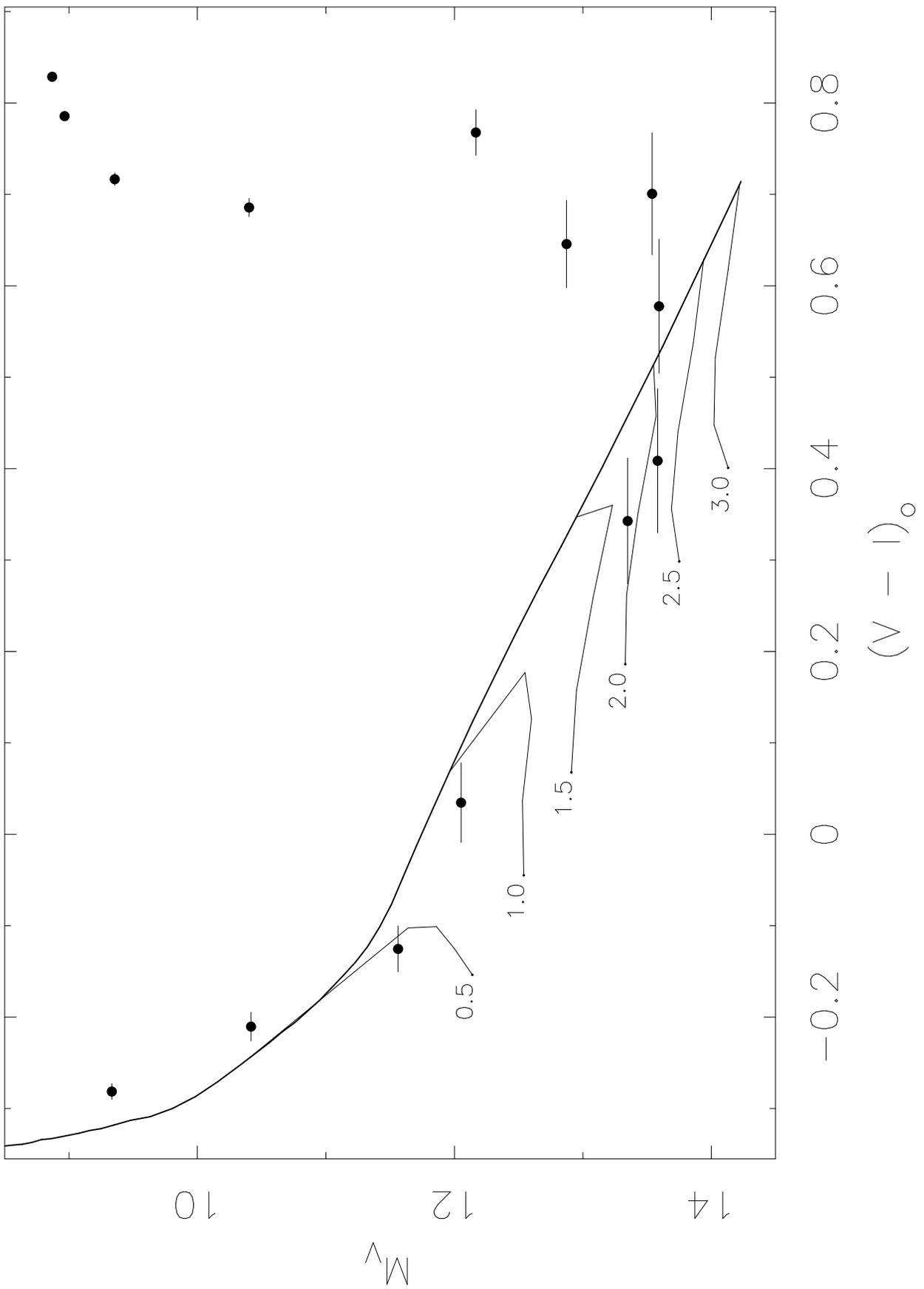
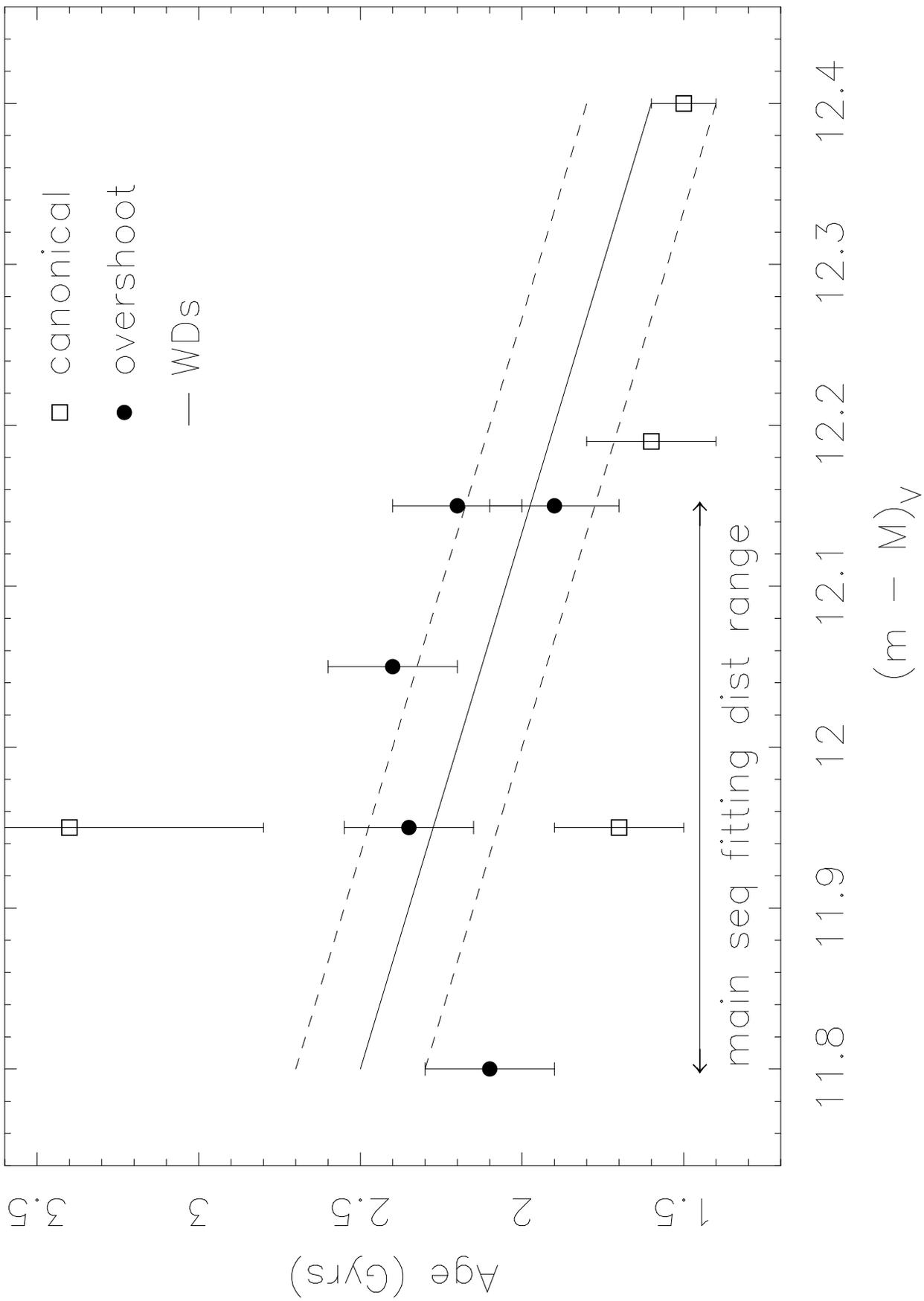
\end{document}